\documentclass[a4paper, 10 pt, conference]{ieeeconf}  

\IEEEoverridecommandlockouts                              
\usepackage{epsfig} 
\usepackage{times} 
\usepackage{amsmath, amsfonts} 
\usepackage{amssymb}  
\usepackage{url}
\usepackage{cite}
\usepackage{hyperref}

\usepackage{algorithmic}
\usepackage{graphicx}
\usepackage{textcomp}
\usepackage{xcolor}

\title{\LARGE \bf
Exact Asymptotic Estimation of Unknown Parameters of Perturbed LRE with Application to State Observation
}

\author{Anton Glushchenko, \textit{Member, IEEE} and Konstantin Lastochkin
\thanks{A. Glushchenko and K. Lastochkin are with V.A. Trapeznikov Institute of Control Sciences of Russian Academy of Sciences, Moscow, Russia
        {\tt\small aiglush@ipu.ru}}%
}

\begin{document}

\maketitle
\thispagestyle{empty}
\pagestyle{empty}

\begin{abstract}
Most identification laws of unknown parameters of linear regression equations (LRE) ensure only boundedness of a parametric error in the presence of additive perturbations, which is almost always unacceptable for practical scenarios. In this paper, a new identification law is proposed to overcome this drawback and guarantee asymptotic convergence of the unknown parameters estimation error to zero in case the mentioned additive perturbation meets special averaging conditions. Such law is successfully applied to state reconstruction problem. Theoretical results are illustrated by numerical simulations. 
\end{abstract}

\section{Preliminaries}

The below-given definitions and notation are used to present the main result of this study.

{\it \bf Definition 1.} \emph{The vector $\varphi \left( t \right) \in {\mathbb{R}^n}$ is finitely exciting \linebreak $(\varphi \in {\rm{FE}})$  over the time range $\left[ {t_r^ + {\rm{,\;}}{t_e}} \right]$  if there exists $t_r^ +  \ge 0$, ${t_e} > t_r^ +$ such that for some $\alpha>0$ the following inequality holds:}
\begin{equation}\label{eq001}
\int\limits_{t_r^ + }^{{t_e}} {\varphi \left( \tau  \right){\varphi ^{\rm{T}}}\left( \tau  \right)d} \tau  \ge \alpha {I_n}{\rm{,}}
\end{equation}
\emph{where $\alpha > 0$ is the excitation level, $I_{n}$ is an identity matrix.}

{\it \bf Definition 2.} \emph{ The vector $ \varphi \left( t \right) \in {\mathbb{R}^n}$ is persistently exciting $\left( {\varphi \in {\rm{PE}}} \right)$ if $\exists T > 0$ and $\alpha  > 0$ such that $\forall t \ge \linebreak \ge {t_0} \ge 0$ the following inequality holds:}
\begin{gather} \label{eq002}
\int\limits_t^{t + T} { \varphi \left( \tau  \right){{ \varphi }^{\rm{T}}}\left( \tau  \right)d} \tau  \ge \alpha {I_n}.
\end{gather}

\textbf{Notation.} Further the following notation is used: $\left| . \right|$ is the absolute value, ${I_{n \times n}}=I_{n}$ is an identity $n \times n$ matrix, ${0_{n \times n}}$ is a zero $n \times n$ matrix, $0_{n}$ stands for a zero vector of length $n$, ${\rm{det}}\{.\}$ stands for a matrix determinant, ${\rm{adj}}\{.\}$ represents an adjoint matrix. We also use the fact that for all (possibly singular) ${n \times n}$ matrices $M$ the following holds: ${\rm{adj}} \{M\} M  =\linebreak= {\rm{det}} \{M\}I_{n \times n}.$

\section{Introduction}
\label{sec:introduction}
The problems of unmeasured state estimation and adaptive control \cite{b1} are often reduced to the one of unknown parameters estimation of LRE:
\begin{equation}\label{eq1}
y\left( t \right) = {\varphi ^{\rm{T}}}\left( t \right)\theta  + w\left( t \right){\rm{,}}
\end{equation}
where $y\left( t \right) \in \mathbb{R}$ is a measurable output signal, $\varphi \left( t \right) \in {\mathbb{R}^n}$ stands for a bounded known regressor, $w\left( t \right) \in \mathbb{R}$ denotes an additive perturbation and $\theta  \in {\mathbb{R}^n}$ is a vector of unknown constant parameters. 

The classical methods to solve the problem of regression \eqref{eq1} parameter estimation are the direct and recursive least squares approaches, as well as the gradient identification law \cite{b1}. In recent years, the procedure of dynamic regressor extension and mixing (DREM) has attracted great attention \cite{b2, b3, b4, b5} as, {in comparison with classical approaches \cite{b1},} it allows one to design the estimation law with relaxed convergence conditions and provides an improved transient quality of the parameter estimates. 

The DREM procedure consists of two main stages: dynamic extension and mixing. In the first step, the original equation \eqref{eq1} is transformed by linear operations to a regression equation with a new regressor, which is a square positive semidefinite matrix. In the second step, an algebraic transformation is applied to reduce the equation obtained at the extension stage into a set of scalar independent equations with respect to the components of the vector $\theta$. Then, the separable estimation laws are introduced.

One of the existing regressor extension schemes (alternative ones are thoroughly described in \cite{b3, b4, b5}) is the Kreisselmeier one proposed in \cite{b6}, according to which the extended vector $Y\left( t \right) \in {{{\mathbb{R}}} ^n}$ and matrix $\Phi \left( t \right) \in {\mathbb{R}^{n \times n}}$ that satisfy a new set of equations:
\begin{equation}\label{eq2}
Y\left( t \right) = \Phi \left( t \right)\theta  + W\left( t \right){\rm{,}}
\end{equation}
are formed as solutions of the following equations:
\begin{equation}\label{eq3}
\begin{array}{l}
\dot Y\left( t \right) =  - lY\left( t \right) + \varphi \left( t \right)y\left( t \right){\rm{,\;}}Y\left( {{t_0}} \right) = {0_n}{\rm{,}}\\
\dot \Phi \left( t \right) =  - l\Phi \left( t \right) + \varphi \left( t \right){\varphi ^{\rm{T}}}\left( t \right){\rm{,\;}}\Phi \left( {{t_0}} \right) = {0_{n \times n}}{\rm{,}}
\end{array}
\end{equation}
and new disturbance $W\left( t \right)$ meets the equation:
\begin{equation}\label{eq4}
\dot W\left( t \right) =  - lW\left( t \right) + \varphi \left( t \right)w\left( t \right){\rm{,\;}}W\left( {{t_0}} \right) = {0_n}.
\end{equation}

Considering the mixing step, the left- and right-hand parts of equation \eqref{eq2} are multiplied by an adjoint matrix ${\rm{adj}}\left\{ {\Phi \left( t \right)} \right\}$, which, owing to ${\rm{adj}}\left\{ {\Phi \left( t \right)} \right\}\Phi \left( t \right) = {\rm{det}}\left\{ {\Phi \left( t \right)} \right\}{I_{n \times n}}$, allows one to obtain a set of scalar regression equations: 
\begin{equation}\label{eq5}
{{\cal Y}_i}\left( t \right) = \Delta \left( t \right){\theta _i} + {{\cal W}_i}\left( t \right){\rm{,}}
\end{equation}
where
\begin{equation*}
\begin{array}{c}
{\cal Y}\left( t \right){\rm{:}} = {\rm{adj}}\left\{ {\Phi \left( t \right)} \right\}Y\left( t \right){\rm{,\quad\quad}}\Delta \left( t \right){\rm{:}} = {\rm{det}}\left\{ {\Phi \left( t \right)} \right\}{\rm{,}}\\
{\cal W}\left( t \right){\rm{:}} = {\rm{adj}}\left\{ {\Phi \left( t \right)} \right\}W\left( t \right){\rm{,}}\\
{\cal Y}\left( t \right) = {{\begin{bmatrix}
{{{\cal Y}_1}\left( t \right)}& \ldots &{{{\cal Y}_{i - 1}}\left( t \right)}&{\begin{array}{*{20}{c}}
 \ldots &{{{\cal Y}_n}\left( t \right)}
\end{array}}
\end{bmatrix}}^{\rm{T}}}{\rm{,}}\\
{\cal W}\left( t \right) = {{\begin{bmatrix}
{{{\cal W}_1}\left( t \right)}& \ldots &{{{\cal W}_{i - 1}}\left( t \right)}&{\begin{array}{*{20}{c}}
 \ldots &{{{\cal W}_n}\left( t \right)}
\end{array}}
\end{bmatrix}}^{\rm{T}}}.
\end{array}
\end{equation*}

Using \eqref{eq5}, the gradient estimation law is introduced:
\begin{equation}\label{eq6}
\begin{array}{l}
\dot {\hat \theta} \left( t \right) = \dot {\tilde {\theta}} \left( t \right) =  - \gamma \Delta \left( t \right)\left( {\Delta \left( t \right)\hat \theta \left( t \right) - {\cal Y}\left( t \right)} \right) = \\
 =  - \gamma {\Delta ^2}\left( t \right)\tilde \theta \left( t \right) + \gamma \Delta \left( t \right){\cal W}\left( t \right){\rm{,\;}}\tilde \theta \left( {{t_0}} \right) = {{\tilde \theta }_0}{\rm{,}}
\end{array}
\end{equation}
where $\gamma > 0$.

An important feature of the extension scheme \eqref{eq3} is that it {\textit{bona fide}} propagates the excitation of the initial regressor $\varphi \left( t \right)$ through all stages of DREM. Thus in \cite{b5} it has been proved that:
\begin{equation}\label{eq7}
\begin{array}{c}
\varphi  \in {\rm{PE}} \Leftrightarrow \Delta  \in {\rm{PE}}\\
 \Updownarrow \\
\exists T > {t_0}{\rm{\;}}\forall t \ge T{\rm{\;}}\Delta \left( t \right) \ge {\Delta _{{\rm{LB}}}} > 0.
\end{array}
\end{equation}

The properties of the estimation error $\tilde \theta \left( t \right)$ when the excitation is propagated and an external perturbation affects the LRE \eqref{eq1} have been studied in \cite{b7, b8, b9} and can be summarized as follows ($c_{\cal{W}}$ is an upper bound of an appropriate norm):
\begin{enumerate}
\item [\textbf{P1.}] if ${{\cal W}_i}\in L_{1}$ and $\Delta  \in {\rm{PE}}$ or $\Delta \notin {L_2}$ then 
\begin{equation*}
\mathop {{\rm{lim}}}\limits_{t \to \infty } \left| {{{\tilde \theta }_i}\left( t \right)} \right| = 0.\end{equation*}
    \item [\textbf{P2.}]if ${{\cal W}_i}\in L_{2}$ and $\Delta  \in {\rm{PE}}$ or $\Delta \notin {L_2}$ then 
    \begin{equation*}
        \mathop {{\rm{lim}}}\limits_{t \to \infty } \left| {{{\tilde \theta }_i}\left( t \right)} \right| \le {c_{\cal W}}\sqrt {{\textstyle{\gamma  \over 2}}}.
        \end{equation*}
            \item [\textbf{P3.}]if ${{\cal W}_i}\in L_{\infty}$ and $\Delta  \in {\rm{PE}}$ or $\Delta \notin {L_2}$ then 
            \begin{equation*}
                \mathop {{\rm{lim}}}\limits_{t \to \infty } \left| {{{\tilde \theta }_i}\left( t \right)} \right| \le {\textstyle{{{c_{\cal W}}{\Delta _{{\rm{UB}}}}} \over {\Delta _{{\rm{LB}}}^2}}}.
                \end{equation*}
\end{enumerate}

The proof of \textbf{P1} is given in \cite[Lemma 1]{b7}, \cite[Lemma 3.1]{b8}, the proof of \textbf{P2} is presented in \cite[Proposition 1]{b9}, and the proof of \textbf{P3} is trivial. Thus \textbf{P1-P3} illustrate that, using the law \eqref{eq6}, the accurate asymptotic estimates of the unknown parameters $\theta$ of equation \eqref{eq1} can be obtained: \linebreak (\textit{i}) in the disturbance-free case or (\textit{ii}) in case of integrable perturbations. This result is almost always impractical and motivates one to state the following problem.

\section{Problem Statement}

Assume that the disturbance ${{\cal W}_i}\left( t \right)$ satisfies the following \emph{averaging} conditions (${{\cal W}_{{\rm{max}}}}$ and ${c_{\cal W}}$ are unknown): {
\begin{enumerate}
    \item [\textbf{C1.}] $\left| {{{\cal W}_i}\left( t \right)} \right| \le {{\cal W}_{{\rm{max}}}}$;
    \item [\textbf{C2.}] $\left| {\int\limits_{{t_0}}^t {\Delta^{-1}(s){\cal{W}}_{i}(s)ds} } \right| \le {c_{\cal W}} < \infty$.
\end{enumerate}}

The aim is to design an estimation law, which, when conditions \textbf{C1-C2} are met, ensures that:
\begin{equation}\label{eq8}
\mathop {{\rm{lim}}}\limits_{t \to \infty } \left| {{{\tilde \theta }_i}\left( t \right)} \right| = {\rm{0}}{\rm{.}}
\end{equation}

{\it \bf Remark 1.} \emph{The conditions \textbf{C1-C2} mean that ${{\cal W}_i}\left( t \right)$ and its weighted primitive function are bounded from above. {It should be specialy noted that meeting \textbf{C2} condition in the general case does not imply that ${{\cal W}_i} \in L_{1}$.}}

{\it \bf Remark 2.} \emph{It can be easily seen from the Fourier analysis  that the requirement \textbf{C2} is met when the perturbation ${{\cal W}_i}\left( t \right)$ and regressor $\Delta\left( t \right)$ do not have common frequncies. For example, when ${{\cal W}_i}\left( t \right) = A{\rm{sin}}\left({\omega}t+\phi\right)$ and $\Delta\left(t\right)=1$, the condition \textbf{C2} is met for any $\omega>0$.}

\section{Main Result}

The main result is described by the following theorem.

{\it \bf Theorem 1.} \emph{Define the estimation law with averaging as follows:}
\begin{equation}\label{eq9}
\begin{array}{l}
{{{\dot {\hat \theta} }_i}\left( t \right) =  - \frac{1}{{{F_i}\left( t \right)}}\left( {{{\hat \theta }_i}\left( t \right) - {\vartheta _i}\left( t \right)} \right){\rm{,\;}}{{\hat \theta }_i}\left( {{t_0}} \right) = {{\hat \theta }_{0i}},}\\
{{{\vartheta }_i}\left( t \right)} = \hat \kappa \left( t \right){{\cal Y}_i}\left( t \right),\\
\dot {\hat \kappa} \left( t \right) =  - \gamma \Delta \left( t \right)\left( {\Delta \left( t \right)\hat \kappa \left( t \right) - 1} \right) - \\ - \dot \Delta \left( t \right){{\hat \kappa }^2}\left( t \right){\rm{,\;}}\hat \kappa \left( {{t_0}} \right) = {{\hat \kappa }_0},\\
\dot \Delta \left( t \right) = {\rm{tr}}\left( {{\rm{adj}}\left\{ {\Phi \left( t \right)} \right\}\dot \Phi \left( t \right)} \right){\rm{,\;}}\Delta \left( {{t_0}} \right) = 0.
\end{array}
\end{equation}
{\emph{where ${F_i}\left( t \right) = t + k_{i}{\rm{,\;}}k_{i} > 0$}.}

\emph{Then, if $\Delta  \in {\rm{PE}}$ and $\gamma > 0$ are chosen such that there exists $\eta  > 0$ satisfying the \textbf{verifiable} inequality:}
\begin{equation*}
\gamma {\Delta ^3}\left( t \right) + \Delta \left( t \right)\dot \Delta \left( t \right)\hat \kappa \left( t \right) + \dot \Delta \left( t \right) \ge \eta \Delta \left( t \right) > 0{\rm{\;}}\forall t \ge T {\rm ,}
\end{equation*}
\emph{the following statements hold:}
\begin{enumerate}
    \item [\textbf{S1)}] \emph{if} \textbf{C1} \emph{is met, then ${{{\tilde \theta}_i}}\in L_{\infty}$ and}
    {\begin{equation*}
        \!\!\!\!\!\mathop {{\rm{lim}}}\limits_{t \to \infty } \left| {{{\tilde \theta }_i}\left( t \right)} \right| \!\le\! \left( {\left| {\tilde \kappa \left( {{t_0}} \right)} \right| \!+\! \Delta _{{\rm{LB}}}^{ - 1}} \right){{\cal W}_{{\rm{max}}}} + {\Delta _{{\rm{UB}}}}\left| {{\theta _i}} \right|\left| {\tilde \kappa \left( {{t_0}} \right)} \right|,\end{equation*}}
    \item [\textbf{S2)}] \emph{if} \textbf{C1} \emph{and} \textbf{C2} \emph{are satisfied, then goal \eqref{eq8} is achieved.}
\end{enumerate}

\emph{Proof of Theorem 1 is postponed to Appendix.}

~

The main concept of the estimation law \eqref{eq9} is as follows. When $\Delta  \in {\rm{PE}}$ and $\gamma > 0$ is large enough, then it is ensured that $\hat \kappa \left( t \right) \to {\Delta ^{ - 1}}\left( t \right)$ as $t \to \infty $ (with exponential rate), and, as a consequence, it can be approximately written:
\begin{equation*}
\begin{array}{l}
{\hat \theta _i}\left( t \right) \approx \frac{1}{{{F_i}\left( t \right)}}\int\limits_{{t_0}}^t {{\theta _i}ds}  + \frac{1}{{{F_i}\left( t \right)}}\left[ {{{\hat \theta }_{0i}} + \int\limits_{{t_0}}^t {{\Delta ^{ - 1}}\left( s \right){{\cal W}_i}\left( s \right)ds} } \right],
\end{array}
\end{equation*}
then, owing to the definition of $F_{i}(t)$ and the facts that the disturbance ${{\cal W}_i}\left( t \right)$ meets averaging conditions \textbf{C1-C2}, the following limit also holds:
{\begin{equation*}
\begin{array}{l}
\mathop {{\rm{lim}}}\limits_{t \to \infty } \frac{1}{{{F_i}\left( t \right)}}\left| {\int\limits_{{t_0}}^t {{\Delta ^{ - 1}}\left( s \right){{\cal W}_i}\left( s \right)ds} } \right| \le \mathop {{\rm{lim}}}\limits_{t \to \infty } \frac{1}{{{F_i}\left( t \right)}}{c_{\cal W}} = 0,\\
\mathop {{\rm{lim}}}\limits_{t \to \infty } \frac{{{{\hat \theta }_{0i}}}}{{{F_i}\left( t \right)}} = 0,{\rm{ }}\mathop {{\rm{lim}}}\limits_{t \to \infty } \frac{1}{{{F_i}\left( t \right)}}\int\limits_{{t_0}}^t {ds}  = {\rm{1}}{\rm{,}}
\end{array}
\end{equation*}}
which allows one to achieve the stated goal \eqref{eq8}.

The proposed estimation law \eqref{eq9} can be easily combined with known identification procedures with relaxed regressor excitation requirements. For example, the combination of \eqref{eq9} with the regressor extension procedure from \cite[Proposition 2]{b4} allows one to ensure that \eqref{eq8} holds under necessary and sufficient \cite[Proposition 1]{b4} identifiability conditions.

{\it \bf Theorem 2.} \emph{Let $\varphi  \in {\rm{FE}}$, then the estimation law with averaging \eqref{eq9}, for which the signals ${\cal Y}\left( t \right){\rm{,\;}}{\cal W}\left( t \right)$ and $\Delta \left( t \right)$ are calculated as follows:}
\begin{equation}\label{eq13}
\begin{array}{c}
{\cal Y}\left( t \right){\rm{:}} = {\rm{adj}}\left\{ {{I_n} - \Phi \left( {t{\rm{,\;}}{t_0}} \right)} \right\}Y\left( t \right){\rm{,}}\\\Delta \left( t \right){\rm{:}} = {\rm{det}}\left\{ {{I_n} - \Phi \left( {t{\rm{,\;}}{t_0}} \right)} \right\}{\rm{,}}\\
{\cal W}\left( t \right){\rm{:}} = {\rm{adj}}\left\{ {{I_n} - \Phi \left( {t{\rm{,\;}}{t_0}} \right)} \right\}W\left( t \right){\rm{,}}\\
\end{array}
\end{equation}
\begin{equation*}
\begin{array}{c}
    \dot \Phi \left( {t{\rm{,\;}}{t_0}} \right) =  - \mu \varphi \left( t \right){\varphi ^{\rm{T}}}\left( t \right)\Phi \left( {t{\rm{,\;}}{t_0}} \right){\rm{,\;}}\Phi \left( {t{\rm{,\;}}{t_0}} \right) = {I_n},\\
\dot Y\left( t \right) =  - \mu \varphi \left( t \right)\left( {{\varphi ^{\rm{T}}}\left( t \right)Y\left( t \right) - y\left( t \right)} \right){\rm{,\;}}Y\left( {{t_0}} \right) = {0_n},\\
\dot W\left( t \right) =  - \mu \varphi \left( t \right){\varphi ^{\rm{T}}}\left( t \right)W\left( t \right) + \varphi \left( t \right)w\left( t \right){\rm{,\;}}W\left( t \right) = {0_n}{\rm{ }}
\end{array}
\end{equation*}
\emph{and the value of $\gamma  > 0$ is chosen such that there exists $\eta  > 0$ that satisfies the \textbf{verifiable} inequality:}
\begin{equation*}
\gamma {\Delta ^3}\left( t \right) + \Delta \left( t \right)\dot \Delta \left( t \right)\hat \kappa \left( t \right) + \dot \Delta \left( t \right) \ge \eta \Delta \left( t \right) > 0{\rm{\;}}\forall t \ge T
\end{equation*}
\emph{ensures that \textbf{S1-S2} holds.}

\emph{Proof of Theorem 2 is presented in Appendix.}

~

To relax the condition $\varphi  \in {\rm{PE}}$ in \eqref{eq13}, the extension scheme proposed in \cite[Proposition 2]{b4} is used instead of the Kreisselmeier’s one \eqref{eq3}. The equations \eqref{eq13} generate a set of scalar regressions \eqref{eq5} and provide $\varphi  \in {\rm{FE}} \Rightarrow \Delta  \in {\rm{PE}}$, which allows one to apply the law \eqref{eq9} and ensure \eqref{eq8} when \textbf{C1-C2} are met.

Thus the proposed identification laws \eqref{eq3} + \eqref{eq9}, \eqref{eq13} + \eqref{eq9} provide estimation error boundedness regardless of the averaging condition \textbf{C2}. If additionally \textbf{C2} is satisfied, the asymptotic convergence of the estimation error to zero is guaranteed.

\section{Application to state observation}

\textcolor{black}{To demonstrate the contribution and significance of the proposed identification law design, we apply it to the task of states estimation of the following nonlinear system:}
\begin{equation}\label{eq13i5}
\begin{array}{l}
\dot x\left( t \right) = Ax\left( t \right) + \phi \left( {y{\rm{,\;}}u} \right) + G\left( {y{\rm{,\;}}u} \right)\theta {\rm{,}}\\
y\left( t \right) = Cx\left( t \right) + \delta \left( t \right){\rm{,\;}}x\left( {{t_0}} \right) = {x_0}{\rm{,}}
\end{array}
\end{equation}
where $x\left( t \right) \in {\mathbb{R}^n}{\rm{,\;}}y\left( t \right) \in {\mathbb{R}^p}{\rm{,\;}}u\left( t \right) \in {\mathbb{R}^m}$ stand for the state, output and control signals, respectively, $\theta  \in {\mathbb{R}^q}$ denotes the vector of unknown parameters, $\delta \left( t \right) \in {\mathbb{R}^p}$ is a bounded exogenous disturbance $\left\| {\delta \left( t \right)} \right\| \le {\delta _{{\rm{max}}}}$, the mappings $\phi {\rm{:\;}}{\mathbb{R}^p} \times {\mathbb{R}^m} \mapsto {\mathbb{R}^n}$ and $G{\rm{:\;}}{\mathbb{R}^p} \times {\mathbb{R}^m} \mapsto {\mathbb{R}^{n \times q}}$ are known and ensure existence and continuability of solutions of the system \eqref{eq13i5}. The matrices $A \in {\mathbb{R}^{n \times n}}{\rm{,\;}}{{C}} \in {\mathbb{R}^{p \times n}}$ are known, and the pair $\left( {A{\rm{,\;}}C} \right)$ is detectable.

The goal is to reconstruct the system state $x\left( t \right)$ such that:
\begin{equation}\label{eq14}
\begin{gathered}
\mathop {{\rm{lim}}}\limits_{t \to \infty } \left\| {\tilde x\left( t \right)} \right\| = \mathop {{\rm{lim}}}\limits_{t \to \infty } \left\| {\hat x\left( t \right) - x\left( t \right)} \right\| \le {\varepsilon _x}\left( {{\delta _{{\rm{max}}}}} \right){\rm{,\;}}\\
\mathop {{\rm{lim}}}\limits_{{\delta _{{\rm{max}}}} \to 0} \left\| {\tilde x\left( t \right)} \right\| = 0.    
\end{gathered}
\end{equation}

The problem under consideration is reduced to the identification of the linear regression equation parameters. For this purpose, the following dynamic filters are introduced:
\begin{equation}\label{eq15}
\begin{array}{l}
\dot \chi \left( t \right) = {A_K}\chi \left( t \right) + Ky\left( t \right){\rm{,\;}}\chi \left( {{t_0}} \right) = {\chi _0}{\rm{,}}\\
\dot P\left( t \right) = {A_K}P\left( t \right) + \phi \left( {y{\rm{,\;}}u} \right){\rm{,\;}}P\left( {{t_0}} \right) = {0_n}{\rm{,}}\\
\dot \Omega \left( t \right) = {A_K}\Omega \left( t \right) + G\left( {y{\rm{,\;}}u} \right){\rm{,\;}}\Omega \left( {{t_0}} \right) = {0_{n \times q}}{\rm{,}}\\
{{\dot \Phi }_K}\left( t \right) = {A_K}{\Phi _K}\left( t \right){\rm{,\;}}{\Phi _K}\left( {{t_0}} \right) = {I_{n \times n}}{\rm{,}}
\end{array}
\end{equation}
where $K \in {\mathbb{R}^{n \times p}}$ ensures that the matrix ${A_K} = A - KC$ is Hurwitz one.

Using the filters \eqref{eq15}, the below-given error is defined:
\begin{equation}\label{eq16}
e\left( t \right) = \chi \left( t \right) + P\left( t \right) + \Omega \left( t \right)\theta  - x\left( t \right).
\end{equation}

Owing to equations \eqref{eq13i5} and \eqref{eq15}, the derivative of the error \eqref{eq16} is obtained as:
\begin{equation}\label{eq17}
\begin{array}{l}
\dot e\left( t \right) = {A_K}\chi \left( t \right) + Ky\left( t \right) + {A_K}P\left( t \right) + \phi \left( {y{\rm{,\;}}u} \right) + \\
 + \left[ {{A_K}\Omega \left( t \right) + G\left( {y{\rm{,\;}}u} \right)} \right]\theta  - Ax\left( t \right) - \phi \left( {y{\rm{,\;}}u} \right) - \\ - G\left( {y{\rm{,\;}}u} \right)\theta
 = {A_K}\left( {\chi \left( t \right) \!+\! P\left( t \right) \!+\! \Omega \left( t \right)\theta  \!-\! x\left( t \right)} \right) + \\ + K\delta \left( t \right)
 = {A_K}e\left( t \right) + K\delta \left( t \right){\rm{,\;}}e\left( {{t_0}} \right) = {\chi _0} - {x_0}.
\end{array}
\end{equation}

The solution of the differential equation \eqref{eq17} is written as follows:
\begin{equation}\label{eq18}
e\left( t \right) = {\Phi _K}\left( t \right){e_0} + {\delta _f}\left( t \right){\rm{,}}
\end{equation}
where
\begin{displaymath}
{\dot \delta _f}\left( t \right) = {A_K}{\delta _f}\left( t \right) + K\delta \left( t \right){\rm{,\;}}{\delta _f}\left( {{t_0}} \right) = {0_n}.    
\end{displaymath}

Substitution of \eqref{eq18} into \eqref{eq16} and multiplication of the obtained equation by ${\cal L}C$ (where ${\cal L} = {\begin{bmatrix}
1&1& \cdots &{{1_p}}
\end{bmatrix}}$) yields:
\begin{equation}\label{eq19}
\begin{array}{c}
{\cal L}C\left[ {{\Phi _K}\left( t \right){e_0} + {\delta _f}\left( t \right)} \right] = {\cal L}C\chi \left( t \right) + {\cal L}CP\left( t \right) +\\
+ {\cal L}C\Omega \left( t \right)\theta  - {\cal L}Cx\left( t \right) \Rightarrow \\
z\left( t \right) = {\varphi ^{\rm{T}}}\left( t \right)\theta  + w\left( t \right){\rm{,}}
\end{array}
\end{equation}
where
\begin{displaymath}
\begin{gathered}
  z\left( t \right) = \mathcal{L}\left[ {y\left( t \right) - C\chi \left( t \right) - CP\left( t \right)} \right]{\text{,\;}} \\ 
  {\varphi ^{\text{T}}}\left( t \right) = \mathcal{L}C\Omega \left( t \right){\text{,\;}}w\left( t \right) \!=\!  - \mathcal{L}C{\Phi _K}\left( t \right){e_0} \!-\! \mathcal{L}C{\delta _f}\left( t \right) \!+\! \mathcal{L}\delta . \\ 
\end{gathered}
\end{displaymath}

Application of the dynamic regressor extension and mixing procedure to equation \eqref{eq19} allows one to obtain the regression equation \eqref{eq5}, on the basis of which the proposed unknown parameters identification law \eqref{eq9} can be implemented.

Using the estimates $\hat \theta \left( t \right)$, the state observer is defined as:
\begin{equation}\label{eq20}
\begin{array}{c}
\hat x\left( t \right) = \chi \left( t \right) + P\left( t \right) + \Omega \left( t \right)\hat \theta \left( t \right){\rm{,}}
\end{array}
\end{equation}
then, if \textbf{C1} and \textbf{C2} are met, it is ensured that:
\begin{displaymath}
\begin{gathered}
\mathop {{\rm{lim}}}\limits_{t \to \infty } \left\| {\tilde x\left( t \right)} \right\| = \mathop {{\rm{lim}}}\limits_{t \to \infty } \left\| {\Omega \left( t \right)\tilde \theta \left( t \right) - {\Phi _K}\left( t \right){e_0} + {\delta _f}\left( t \right)} \right\| \le {\varepsilon _x}{\rm{,}}
\end{gathered}
\end{displaymath}
where ${\varepsilon _x} = \left\| {\Pi K} \right\|{\delta _{{\rm{max}}}}\sqrt {{\textstyle{{{c^{ - 1}}{\lambda _{{\rm{max}}}}\left( \Pi  \right)} \over {\left( {{\lambda _{{\rm{min}}}}\left( Q \right) - c} \right){\lambda _{{\rm{min}}}}\left( \Pi  \right)}}}} $, $c > 0$ is a constant such that ${\lambda _{{\rm{min}}}}\left( Q \right) - c > 0$, and $\Pi  = {\Pi ^{\rm{T}}} > 0{\rm{,\;}}Q = {Q^{\rm{T}}} > 0$ meet the Lyapunov equation $A_K^{\rm{T}}\Pi  + \Pi {A_K} =  - Q$.

Hence, when the conditions \textbf{C1} and \textbf{C2} are satisfied, the proposed approach allows one to solve the state estimation problem \eqref{eq14}.

\section{Simulations}

\subsection{Estimation when PE condition is met}

The regression equation \eqref{eq1} was defined as follows:
\begin{equation}\label{eq21}
\begin{array}{l}
\varphi \left( t \right) = {\begin{bmatrix}
{{\rm{sin}}\left( t \right)}\\
1
\end{bmatrix}}{\rm{,\;}}\theta  = {\begin{bmatrix}
1\\
{ - 1}
\end{bmatrix}}{\rm{, }}\\
w\left( t \right) = 1{\rm{sin}}\left( t \right) + 0.2{\rm{sin}}\left( {0.1t} \right).
\end{array}
\end{equation}

The parameters of filters \eqref{eq3} and estimation laws \eqref{eq6}, \eqref{eq9} were chosen as:
\begin{equation*}
l = 1,{\rm{\;}}\gamma  = \left\{ \begin{array}{l}
{\rm{1}}{{\rm{0}}^2}{\rm{,\;for\;}}\eqref{eq6}\\
{\rm{1}}{{\rm{0}}^4}{\rm{,\;for\;}}\eqref{eq9}
\end{array} \right.{\rm{,\;}}k_{i}=10^{-3}{\rm{,\;}}i = \overline {1,{\rm{ 2}}} .
\end{equation*}

Figure 1 depicts the behavior of the regressor $\Delta \left( t \right)$, disturbance ${{\cal W}_i}\left( t \right)$ and comparison of $\gamma {\Delta ^3}\left( t \right) + \Delta \left( t \right)\dot \Delta \left( t \right)\hat \kappa \left( t \right) + \dot \Delta \left( t \right)$ and $\eta \Delta \left( t \right)$ for $\eta  = 50$.

\begin{figure}[thpb]
      \centering
      \includegraphics[scale=0.5]{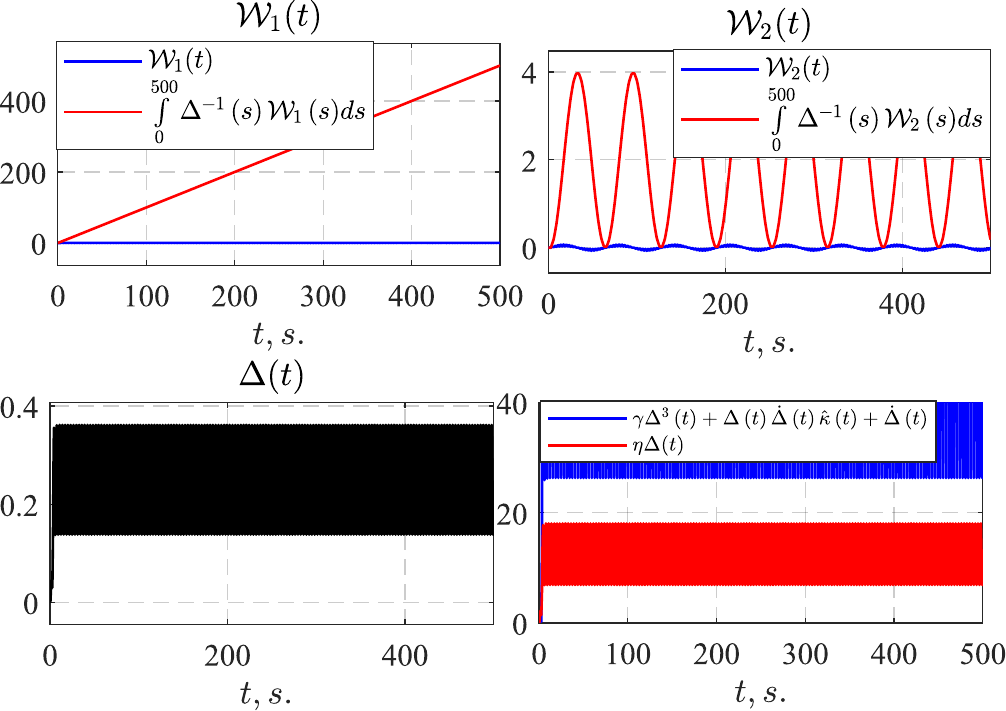}
      \caption{Behavior of $\Delta \left( t \right)$, ${{\cal W}_i}\left( t \right)$ and comparison of $\gamma {\Delta ^3}\left( t \right) + + \Delta \left( t \right)\dot \Delta \left( t \right)\hat \kappa \left( t \right) + \dot \Delta \left( t \right)$ and $\eta \Delta \left( t \right)$ for $\eta  = 50$.}
      \label{Figure1} 
      \end{figure}

      The obtained results allow one to make the following conclusions: 
\begin{enumerate}
\item[\emph{i})] $\Delta  \in {\rm{PE}}$,
\item[\emph{ii})] there exists $\eta  = 50$ such that
\begin{equation*}
\gamma {\Delta ^3}\left( t \right) + \Delta \left( t \right)\dot \Delta \left( t \right)\hat \kappa \left( t \right) + \dot \Delta \left( t \right) \ge \eta \Delta \left( t \right) > 0{\rm{\;}}\forall t \ge T.
\end{equation*}
\item[\emph{iii})]   ${{\cal W}_1}\left( t \right)$ and ${{\cal W}_2}\left( t \right)$ are bounded,
\item[\emph{iv})] ${{\cal W}_2}\left( t \right)$ meets the averaging condition \textbf{C2}.
\end{enumerate}

Figure 2 presents the behavior of  $\hat \theta \left( t \right)$ for \eqref{eq6} and \eqref{eq9}.
\begin{figure}[thpb]
      \centering
      \includegraphics[scale=0.5]{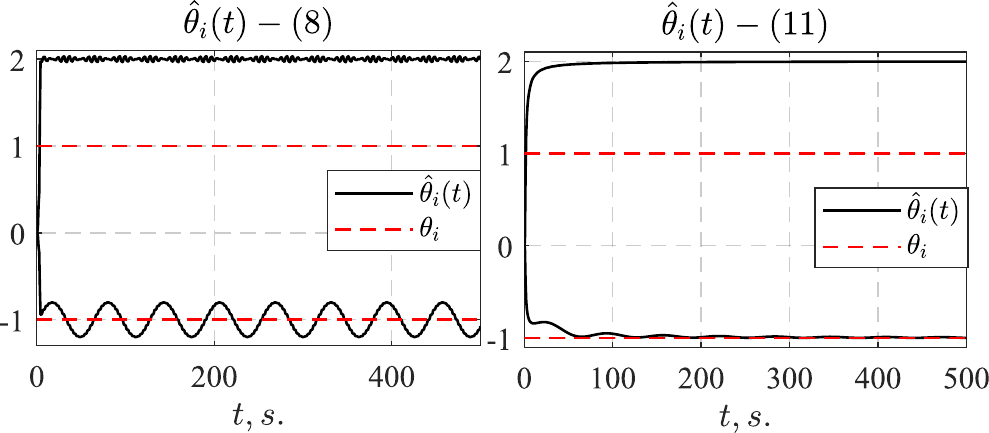}
      \caption{Behavior of  $\hat \theta \left( t \right)$ for \eqref{eq6} and \eqref{eq9}}
      \label{Figure2} 
      \end{figure}

The curves from Fig.1 and Fig.2 validate the theoretical results and show that the proposed estimation law \eqref{eq9}, unlike \eqref{eq6}, ensures asymptotic convergence of the error $\tilde \theta \left( t \right)$ to zero in case the additive perturbation satisfies \textbf{C1-C2} averaging conditions.

\subsection{Estimation when PE condition is violated}

The regression equation \eqref{eq1} was defined as:
\begin{equation}\label{eq22}
\begin{array}{l}
\varphi \left( t \right) = {\begin{bmatrix}
{{e^{ - t}}}\\
1
\end{bmatrix}}{\rm{,\;}}\theta  = {\begin{bmatrix}
1\\
{ - 1}
\end{bmatrix}}{\rm{, }}\\
w\left( t \right) = 1{\rm{sin}}\left( t \right) + 0.2{\rm{sin}}\left( {0.1t} \right){\rm{,}}
\end{array}
\end{equation}

The parameters of filters \eqref{eq13} and estimation laws \eqref{eq13}+\eqref{eq6}, \eqref{eq13}+\eqref{eq9} were picked as:
\begin{equation*}
\gamma  = \left\{ \begin{array}{l}
{\rm{1}}{\rm{,\;for\;}}\eqref{eq13} + \eqref{eq6}\\
250,{\rm{\;for\;}}\eqref{eq13} + \eqref{eq9}
\end{array} \right.{\rm{,\;}}\begin{array}{*{20}{c}}
\mu  = 10\\
k_{i}=10^{-3}
\end{array}{\rm{,\;}}i = \overline {1,{\rm{ 2}}} .
\end{equation*}

Figure 3 depicts the behavior of the regressor $\Delta \left( t \right)$, disturbance ${{\cal W}_i}\left( t \right)$ and comparison of $\gamma {\Delta ^3}\left( t \right) + \Delta \left( t \right)\dot \Delta \left( t \right)\hat \kappa \left( t \right) + \dot \Delta \left( t \right)$ and $\eta \Delta \left( t \right)$ for $\eta  = 10$.

\begin{figure}[thpb]
      \centering
      \includegraphics[scale=0.49]{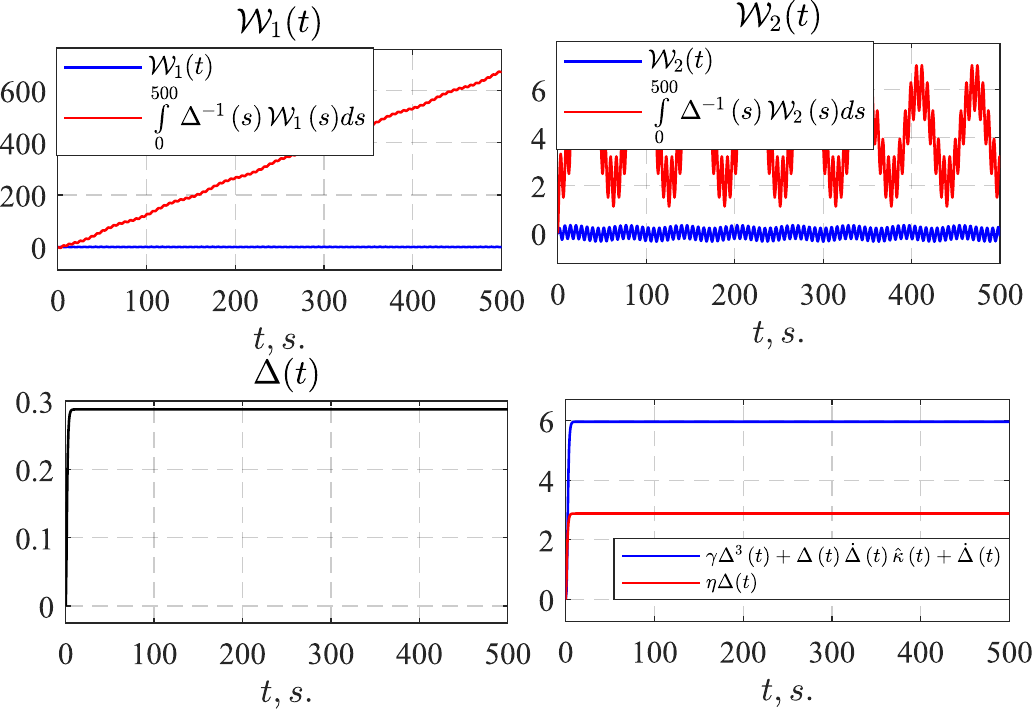}
      \caption{Behavior of $\Delta \left( t \right)$, ${{\cal W}_i}\left( t \right)$ and comparison of $\gamma {\Delta ^3}\left( t \right) + + \Delta \left( t \right)\dot \Delta \left( t \right)\hat \kappa \left( t \right) + \dot \Delta \left( t \right)$ and $\eta \Delta \left( t \right)$ for $\eta  = 10$.}
      \label{Figure3} 
      \end{figure}

      The obtained results allow one to make the following conclusions: 
\begin{enumerate}
\item[\emph{i})] $\Delta  \in {\rm{PE}}$,
\item[\emph{ii})] there exists $\eta  = 10$ such that
\begin{equation*}
\gamma {\Delta ^3}\left( t \right) + \Delta \left( t \right)\dot \Delta \left( t \right)\hat \kappa \left( t \right) + \dot \Delta \left( t \right) \ge \eta \Delta \left( t \right) > 0{\rm{\;}}\forall t \ge T.
\end{equation*}
\item[\emph{iii})]   ${{\cal W}_1}\left( t \right)$ and ${{\cal W}_2}\left( t \right)$ are bounded,
\item[\emph{iv})] ${{\cal W}_2}\left( t \right)$ meets the averaging condition \textbf{C2}.
\end{enumerate}

Figure 4 is to present the behavior of  $\hat \theta \left( t \right)$ for \eqref{eq13} + \eqref{eq6} and \eqref{eq13} + \eqref{eq9}.

\begin{figure}[thpb]
      \centering
      \includegraphics[scale=0.5]{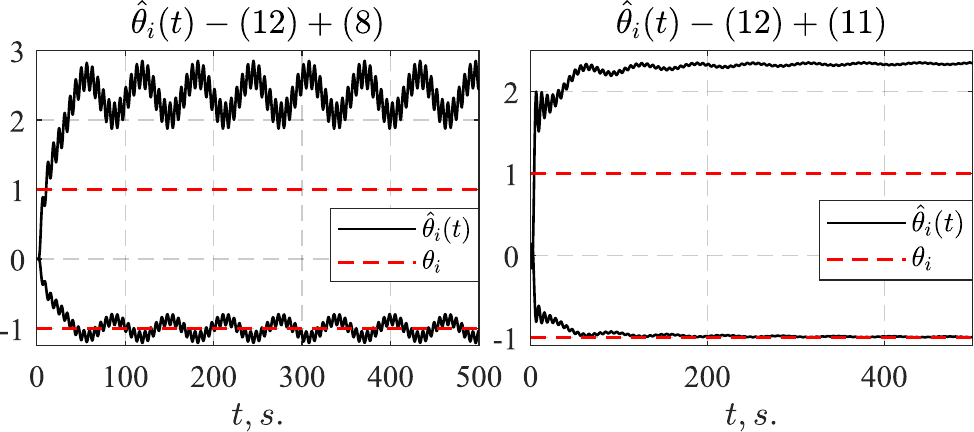}
      \caption{Behavior of $\hat \theta \left( t \right)$ for \eqref{eq13} + \eqref{eq6} and \eqref{eq13} + \eqref{eq9}.}
      \label{Figure4} 
      \end{figure}

      The curves from Fig.3 and Fig.4 validate the theoretical results of Theorem 2, particularly, the error  ${\tilde \theta _1}\left( t \right)$ is bounded and ${\tilde \theta _2}\left( t \right)$ converges asymptotically to zero.

\subsection{State observation example}

A dynamical system with the following parametrization was considered:
\begin{equation}\label{eq23}
\begin{gathered}
A = {\begin{bmatrix}
0&1\\
0&a
\end{bmatrix}} {\rm{,\;}}\phi \left( {y,{\rm{\;}}{u}} \right) =  {\begin{bmatrix}
0\\
{bu}
\end{bmatrix}} {\rm{,\;}}\\
G\left( y,{\rm{\;}}{u} \right) = {\begin{bmatrix}
0\\
b
\end{bmatrix}} \theta {\rm{,\;}}C = {\begin{bmatrix}
1&0
\end{bmatrix}}{\rm{,}}
\end{gathered}
\end{equation}
where
\begin{displaymath}
\begin{array}{c}
a =  - 169.{\rm{5,\;}}b = 16900{\rm{,\;}}\delta \left( t \right) = {\rm{5sin}}\left( {5t} \right) + 0.{\rm{2sin}}\left( t \right) + 1{\rm{,}}\\
u\left( t \right) = 0{\rm{,\;}}\theta  = 0.{\rm{2}}.
\end{array}
\end{displaymath}

The parameters of filters \eqref{eq3}, \eqref{eq15} and estimation laws \eqref{eq6}, \eqref{eq9} were chosen as:
\begin{displaymath}
\begin{array}{c}
l = 1{\rm{,\;}}K = {{\begin{bmatrix}
{100}&1
\end{bmatrix}}^{\rm{T}}}{\rm{,\;}}\\
{k_i} = 0.{\rm{01,\;}}i = \overline {1{\rm{,\;2}}} {\rm{,\;}}\gamma  = \left\{ \begin{array}{l}
{\rm{1}}{{\rm{0}}^2}{\rm{, \; for \;\eqref{eq6}}},\\
{\rm{1}}{{\rm{0}}^3}{\rm{, \; for \; \eqref{eq9}}}.
\end{array} \right.
\end{array}
\end{displaymath}

Figure 5 depicts the behavior of the regressor $\Delta \left( t \right)$, disturbance ${{\cal W}}\left( t \right)$ and comparison of $\gamma {\Delta ^3}\left( t \right) + \Delta \left( t \right)\dot \Delta \left( t \right)\hat \kappa \left( t \right) + \dot \Delta \left( t \right)$ and $\eta \Delta \left( t \right)$ for $\eta  = 100$.

\begin{figure}[thpb]
      \centering
      \includegraphics[scale=0.5]{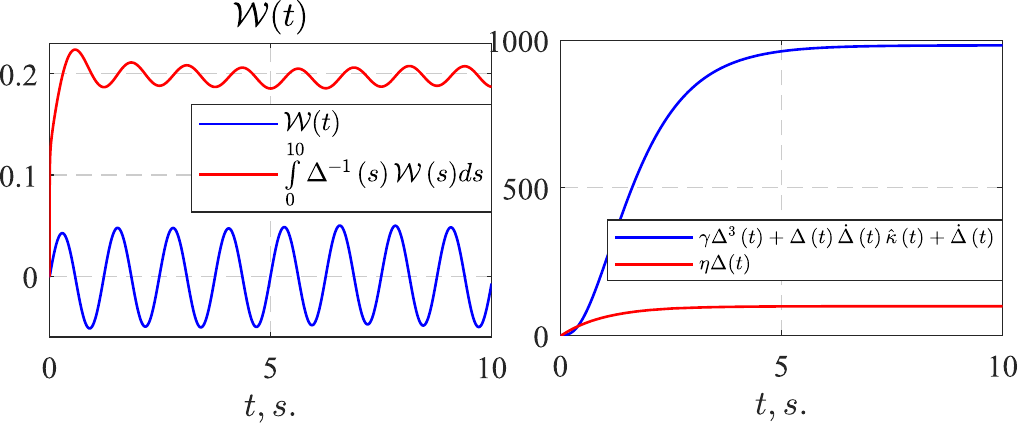}
      \caption{Behavior of $\Delta \left( t \right)$, ${{\cal W}_i}\left( t \right)$ and comparison of $\gamma {\Delta ^3}\left( t \right) + + \Delta \left( t \right)\dot \Delta \left( t \right)\hat \kappa \left( t \right) + \dot \Delta \left( t \right)$ and $\eta \Delta \left( t \right)$ for $\eta  = 100$.}
      \label{Figure1} 
      \end{figure}

The obtained results allow one to make the following conclusions: 
\begin{enumerate}
\item[\emph{i})] $\Delta  \in {\rm{PE}}$,
\item[\emph{ii})] there exists $\eta  = 100$ such that
\begin{equation*}
\!\!\!\!\!\gamma {\Delta ^3}\left( t \right) + \Delta \left( t \right)\dot \Delta \left( t \right)\hat \kappa \left( t \right) + \dot \Delta \left( t \right) \ge \eta \Delta \left( t \right) > 0{\rm{\;}}\forall t \ge T.
\end{equation*}
\item[\emph{iii})]   ${{\cal W}}\left( t \right)$ is bounded,
\item[\emph{iv})] ${{\cal W}}\left( t \right)$ meets the averaging condition \textbf{C2}.
\end{enumerate}

Figure 6 depicts the behavior of $\hat \theta \left( t \right)$ and $\hat x\left( t \right)$ for \eqref{eq6} and \eqref{eq9}.

\begin{figure}[thpb]
      \centering
      \includegraphics[scale=0.5]{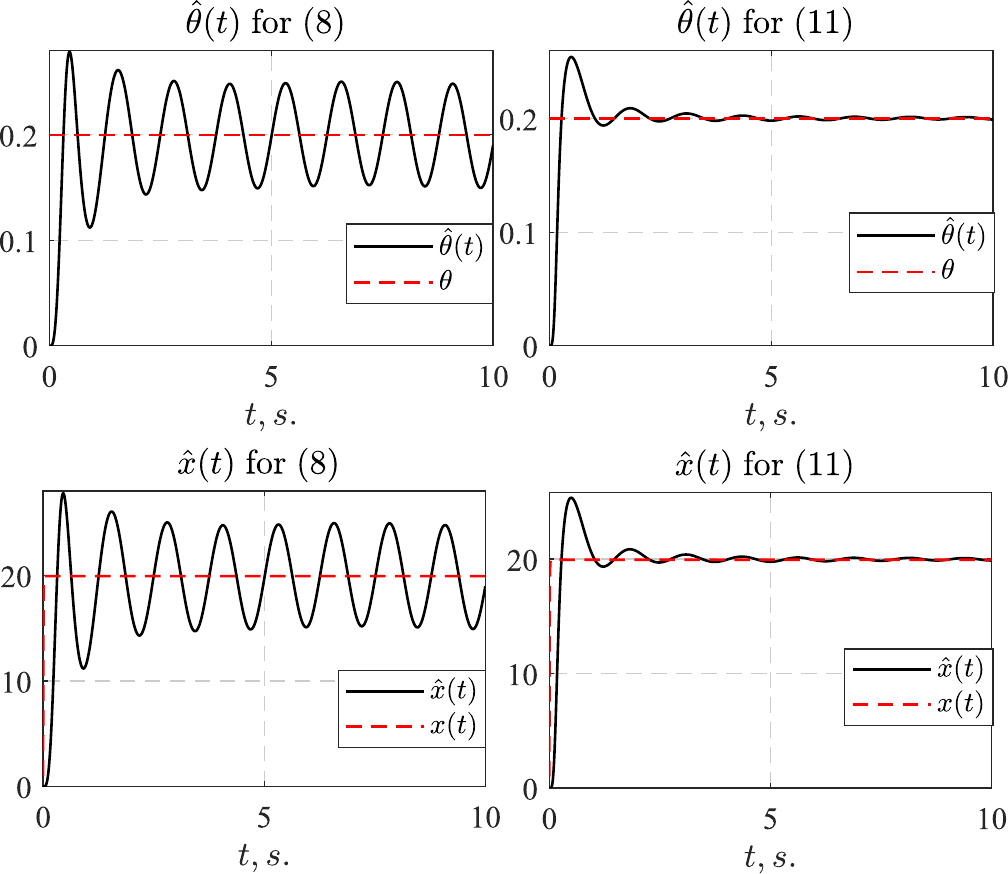}
      \caption{Behavior of $\hat \theta \left( t \right)$ and $\hat x\left( t \right)$ for \eqref{eq6} and \eqref{eq9}.}
      \label{Figure1} 
      \end{figure}

The obtained curves demonstrate the effect of application of the proposed estimation law to the state reconstruction problem. Unlike the algorithm \eqref{eq6}, the steady-state error of unmeasured states reconstruction was significantly reduced.

\section{Conclusion}
The estimation law was proposed that guaranteed asymptotic convergence to zero of the estimation error of the scalar regression equation unknown parameters when the disturbance met averaging conditions \textbf{C1-C2}. This estimation law can be used to solve any problems of adaptive control theory that are reducible to the one of equation \eqref{eq1} parameters identification. The scope of further research is to increase the rate of convergence of parameter estimates to their true values.


\appendices

\renewcommand{\theequation}{A\arabic{equation}}
\setcounter{equation}{0}  

\section*{Appendix}
{\it Proof of Theorem 1.} For all $t \ge T$ define the following error:
\begin{equation*} 
\tilde \kappa \left( t \right) = \hat \kappa \left( t \right) - {\Delta ^{ - 1}}\left( t \right){\rm{,}}
\end{equation*}
which is differentiated and consequently, owing to 
\begin{equation*} 
\begin{array}{c}
\Delta \left( t \right){\Delta ^{ - 1}}\left( t \right) = 1 \Leftrightarrow \dot \Delta \left( t \right){\Delta ^{ - 1}}\left( t \right) + \Delta \left( t \right){\textstyle{{d{\Delta ^{ - 1}}\left( t \right)} \over {dt}}} = 0\\
 \Updownarrow \\
{\textstyle{{d{\Delta ^{ - 1}}\left( t \right)} \over {dt}}} =  - \dot \Delta \left( t \right){\Delta ^{ - 2}}\left( t \right),
\end{array}
\end{equation*}
it is obtained: 
\begin{equation} {\label{eqA2}}
\begin{array}{l}
\dot {\tilde \kappa}  =  - \gamma \Delta \left( {\Delta \hat \kappa  - 1} \right) - \dot \Delta {{\hat \kappa }^2} + \dot \Delta {\Delta ^{ - 2}} = \\
 =  - \gamma {\Delta ^2}\tilde \kappa  - \dot \Delta \left( {\hat \kappa  + {\Delta ^{ - 1}}} \right)\tilde \kappa = \\
 =  - \left( {\gamma {\Delta ^2} + \dot \Delta \hat \kappa  + \dot \Delta {\Delta ^{ - 1}}} \right)\tilde \kappa {\rm{,}}
\end{array}
\end{equation}
where $\dot \Delta \left( t \right)$ obeys Jacobi’s formula (see Theorem 8.1 in \cite{b10}):
\begin{equation*} 
\dot \Delta \left( t \right) = {\rm{tr}}\left( {{\rm{adj}}\left\{ {\Phi \left( t \right)} \right\}\dot \Phi \left( t \right)} \right){\rm{,\;}}\Delta \left( {{t_0}} \right) = 0.
\end{equation*}

The quadratic form $V\left( t \right) = {\textstyle{1 \over 2}}{\tilde \kappa ^2}\left( t \right)$ is introduced, which derivative is written as:
\begin{equation*} 
\dot V\left( t \right) =  - 2\left( {\gamma {\Delta ^2}\left( t \right) + \dot \Delta \left( t \right)\hat \kappa \left( t \right) + \dot \Delta \left( t \right){\Delta ^{ - 1}}\left( t \right)} \right)V\left( t \right){\rm{,}}
\end{equation*}
from which, when $\gamma {\Delta ^3}\left( t \right) + \Delta \left( t \right)\dot \Delta \left( t \right)\hat \kappa \left( t \right) + \dot \Delta \left( t \right) \ge \eta \Delta \left( t \right) > 0$, then for all $t \ge T$ there exists the following upper bound
\begin{equation} {\label{eqA3}}
\left| {\tilde \kappa \left( t \right)} \right| \le {e^{ - \eta \left( {t - T} \right)}}\left| {\tilde \kappa \left( {{t_0}} \right)} \right|
\end{equation}
and, therefore, it holds that: 
\begin{equation} {\label{eqA4}}
\int\limits_T^t {\left| {\tilde \kappa \left( s \right)} \right|ds}  < \infty .
\end{equation}

Equation for ${\vartheta _i}\left( t \right)$ is rewritten in the following form:
\begin{equation} {\label{eqA5}}
\begin{array}{l}
{\vartheta _i}\left( t \right) = \hat \kappa \left( t \right){{\cal Y}_i}\left( t \right) \pm {\Delta ^{ - 1}}\left( t \right){{\cal Y}_i}\left( t \right) =\\={\Delta ^{ - 1}}\left( t \right){{\cal Y}_i}\left( t \right) + \tilde \kappa \left( t \right){{\cal Y}_i}\left( t \right).
\end{array}
\end{equation}

Let’s now introduce the following function:
\begin{equation} {\label{eqA6}}
L\left( t \right) = F\left( t \right){\tilde \theta _i}\left( t \right).
\end{equation}

Then, we differentiate \eqref{eqA6} with respect to time and substitute equations for ${\dot{ \hat \theta} _i}\left( t \right)$ and ${\vartheta _i}\left( t \right)$ into the obtained result:
\begin{equation} {\label{eqA7}}
\begin{array}{l}
\dot L\left( t \right) = {\textstyle{{dF\left( t \right)} \over {dt}}}{{\tilde \theta }_i}\left( t \right) + F\left( t \right){{\dot {\tilde \theta} }_i}\left( t \right) = \\ = {\textstyle{{dF\left( t \right)} \over {dt}}}{{\tilde \theta }_i}\left( t \right) - \left( {{{\hat \theta }_i}\left( t \right) \pm {\theta _i} - {\vartheta _i}\left( t \right)} \right) = \\
 = \left( {{\textstyle{{dF\left( t \right)} \over {dt}}} - 1} \right){{\tilde \theta }_i}\left( t \right) - \\- {\theta _i} + {\Delta ^{ - 1}}\left( t \right){{\cal Y}_i}\left( t \right) + \tilde \kappa \left( t \right){{\cal Y}_i}\left( t \right) = \\
 =  - {\theta _i} + {\Delta ^{ - 1}}\left( t \right){{\cal Y}_i}\left( t \right) + \tilde \kappa \left( t \right){{\cal Y}_i}\left( t \right).
\end{array}
\end{equation}

The solution of \eqref{eqA7} is obtained as:
\begin{equation} {\label{eqA8}}
\begin{array}{l}
{{\tilde \theta }_i}\left( t \right) = \\ = \frac{1}{{{F_i}\left( t \right)}}\left[ {\int\limits_{{t_0}}^T {\dot L\left( s \right)} ds + \int\limits_T^t {\dot L\left( s \right)} ds} \right] + {\textstyle{{{F_i}\left( {{t_0}} \right){{\tilde \theta }_i}\left( {{t_0}} \right)} \over {{F_i}\left( t \right)}}} = \\
 = \frac{1}{{{F_i}\left( t \right)}}\int\limits_T^t {{\Delta ^{ - 1}}\left( s \right){{\cal W}_i}\left( s \right) \!+\! \tilde \kappa \left( s \right){{\cal Y}_i}\left( s \right)} ds + {\textstyle{{{{\tilde \theta }_i}\left( T \right)} \over {{F_i}\left( t \right)}}}{\rm{,}}
\end{array}
\end{equation}
where, as $0 \le \Delta \left( t \right) \le {\Delta _{{\rm{LB}}}}$ for all $t \in \left[ {{t_0}{\rm{,\;}}{{ T}} } \right)$ and $\tilde \kappa \left( t \right)$ is continuous (does not have finite escape time), it is assumed that sum ${F_i}\left( {{t_0}} \right){\tilde \theta _i}\left( {{t_0}} \right) + \int\limits_{{t_0}}^T {\dot L\left( s \right)} ds$ has some finite value ${\tilde \theta _i}\left( T \right)$.

\textbf{S1)} If $\left| {{{\cal W}_i}\left( t \right)} \right| \le {{\cal W}_{{\rm{max}}}} < \infty $, then, considering \eqref{eqA8}, the following bound of ${\tilde \theta _i}\left( t \right)$ is written:
\begin{equation} {\label{eqA9}}
\begin{array}{l}
\left| {{{\tilde \theta }_i}\left( t \right)} \right| \le \left| {\frac{1}{{{F_i}\left( t \right)}}} \right|\left( {\left| {{{\tilde \theta }_i}\left( T \right)} \right| + } \right.\\
\left. { + \left| {\int\limits_T^t {{\Delta ^{ - 1}}\left( s \right){{\cal W}_i}\left( s \right)ds}  + \int\limits_T^t {\tilde \kappa \left( s \right){{\cal Y}_i}\left( s \right)ds} } \right|} \right) \le \\
 \le \left| {\frac{1}{{{F_i}\left( t \right)}}} \right|\left( {{{\tilde \theta }_i}\left( T \right) + \int\limits_T^t {ds} \Delta _{{\rm{LB}}}^{ - 1}{{\cal W}_{{\rm{max}}}} + } \right.\\
\left. { + \left| {\int\limits_T^t {\tilde \kappa \left( s \right)\Delta \left( s \right){\theta _i}ds} } \right| + \left| {\int\limits_T^t {\tilde \kappa \left( s \right){{\cal W}_i}\left( s \right)ds} } \right|} \right) \le \\
 \le \left| {\frac{1}{{{F_i}\left( t \right)}}} \right|\left( {{{\tilde \theta }_i}\left( T \right) + \int\limits_T^t {ds} \left( {\Delta _{{\rm{LB}}}^{ - 1} \!+\! \left| {\tilde \kappa \left( {{t_0}} \right)} \right|} \right){{\cal W}_{{\rm{max}}}} + } \right.\\
\left. { + \int\limits_T^t {ds} {\Delta _{{\rm{UB}}}}\left| {{\theta _i}} \right|\left| {\tilde \kappa \left( {{t_0}} \right)} \right|} \right) < \infty {\rm{,}}
\end{array}
\end{equation}
where $0 < {\Delta _{LB}} \le \Delta \left( t \right) \le {\Delta _{UB}}$ as the regressor $\varphi \left( t \right)$ is bounded and $\Delta  \in {\rm{PE}}$. 

On the basis of \eqref{eqA9} and owing to the definition of $F_{i}(t)$, it holds that $\mathop {{\rm{lim}}}\limits_{t \to \infty } \left| {{{\tilde \theta }_i}\left( t \right)} \right| \le \left( {\left| {\tilde \kappa \left( {{t_0}} \right)} \right| + \Delta _{{\rm{LB}}}^{ - 1}} \right){{\cal W}_{{\rm{max}}}} + {\Delta _{{\rm{UB}}}}\left| {{\theta _i}} \right|\left| {\tilde \kappa \left( {{t_0}} \right)} \right|$, which was to be proved in the first statement of Theorem.

\textbf{S2)} To prove the second statement of Theorem the following upper bound of the error ${\tilde \theta _i}\left( t \right)$ is obtained: 
\begin{equation}
    \begin{array}{l}
\left| {{{\tilde \theta }_i}\left( t \right)} \right| = \left| {\frac{1}{{{F_i}\left( t \right)}}\int\limits_T^t {{\Delta ^{ - 1}}\left( s \right){{\cal W}_i}\left( s \right)ds} } \right| + \\
 + \left| {\frac{1}{{{F_i}\left( t \right)}}\int\limits_T^t {\tilde \kappa \left( s \right){{\cal Y}_i}\left( s \right)ds} } \right| + \left| {\frac{{{{\tilde \theta }_i}\left( T \right)}}{{{F_i}\left( t \right)}}} \right| \le \\
 \le \left| {\frac{1}{{{F_i}\left( t \right)}}{c_{\cal W}}} \right| + \left| {\frac{{{\theta _i}}}{{{F_i}\left( t \right)}}\int\limits_T^t {\tilde \kappa \left( s \right)\Delta \left( s \right)ds} } \right| + \\
  + \left| {\frac{1}{{{F_i}\left( t \right)}}\int\limits_T^t {\tilde \kappa \left( s \right){{\cal W}_i}\left( s \right)ds} } \right| + \left| {\frac{{{{\tilde \theta }_i}\left( T \right)}}{{{F_i}\left( t \right)}}} \right| \le \\
 \le \left| {\frac{1}{{{F_i}\left( t \right)}}{c_{\cal W}}} \right| + \left| {\frac{{{\theta _i}}}{{{F_i}\left( t \right)}}} \right|\int\limits_T^t {\left| {\tilde \kappa \left( s \right)\Delta \left( s \right)} \right|ds}  + \\
 + \left| {\frac{1}{{{F_i}\left( t \right)}}c_{\kappa}} \right| + \left| {\frac{{{{\tilde \theta }_i}\left( T \right)}}{{{F_i}\left( t \right)}}} \right| 
 \le \left| {\frac{1}{{{F_i}\left( t \right)}}\left(c_{\mathcal{W}}+c_{\kappa} \right)} \right|+ \\
 + \left| {\frac{{{\theta _i}}}{{{F_i}\left( t \right)}}} \right|\underbrace {\int\limits_T^t {\left| {\tilde \kappa \left( s \right)} \right|ds} }_{ < \infty }{\Delta _{{\rm{UB}}}} + \left| {\frac{{{{\tilde \theta }_i}\left( T \right)}}{{{F_i}\left( t \right)}}} \right|,
\end{array}
 {\normalsize\label{eqA10}}
\end{equation}
where $0 < {\Delta _{\rm{LB}}} \le \Delta \left( t \right) \le {\Delta _{\rm{UB}}}$ as the regressor $\varphi \left( t \right)$ is bounded and $\Delta  \in {\rm{PE}}$, and there exists $c_{\kappa}>0$ such that $\int\limits_T^t {\tilde \kappa \left( s \right){{\cal W}_i}\left( s \right)ds} \le c_{\kappa}<\infty$ as $\mathcal{W}_{i}\left(t\right)$ is bounded from {\textbf{C1}} and $\tilde{\kappa}\left(t\right)$ exponentially converges to zero \eqref{eqA3}.

Owing to definition of $F_{i}(t)$,  it immediately follows from \eqref{eqA10} that the stated goal \eqref{eq8} is achieved. 

{\it Proof of Theorem 2.} The error $\tilde Y\left( t \right) = Y\left( t \right) - \theta$ is introduced, which derivative with respect to time is written as:
\begin{equation}
\begin{array}{c}
\dot {\tilde Y}\left( t \right) =  - \mu \varphi \left( t \right){\varphi ^{\rm{T}}}\left( t \right)\tilde Y\left( t \right) + \mu \varphi \left( t \right)w\left( t \right){\rm{,}}\\\tilde Y\left( {{t_0}} \right) = {Y_0} - \theta .
\end{array}
 {\normalsize\label{eqA11}}
\end{equation}

The solution of the differential equation \eqref{eqA11} is obtained as:
\begin{equation}
\begin{array}{c}
Y\left( t \right) - \theta  = \Phi \left( {t{\rm{,\;}}{t_0}} \right)\left( {Y\left( {{t_0}} \right) - \theta } \right) + W\left( t \right)\\
 \Updownarrow \\
Y\left( t \right) = \left( {{I_n} - \Phi \left( {t{\rm{,\;}}{t_0}} \right)} \right)\theta  + W\left( t \right),
\end{array}
 {\normalsize\label{eqA12}}
\end{equation}
from which, considering \eqref{eq13}, we have:
\begin{equation}
{\cal Y}\left( t \right) = \Delta \left( t \right)\theta  + {\cal W}\left( t \right).
 {\normalsize\label{eqA13}}
\end{equation}

According to the results of Lemma 3 from \cite{b4}, when \linebreak $\varphi  \in {\rm{FE}}$, then $\Delta  \in {\rm{PE}} \Leftrightarrow \exists T > {t_0}{\rm{\;}}\forall t \ge T{\rm{\;}}\Delta \left( t \right) \ge {\Delta _{{\rm{LB}}}} > \linebreak >0$. As $\varphi \left( t \right){\varphi ^{\rm{T}}}\left( t \right)$ is a positive semidefinite matrix, and the regressor $\varphi \left( t \right)$ is bounded, then it holds that $\Delta \left( t \right) \le {\Delta _{UB}}$. Therefore, considering the results from Theorem 1, if \textbf{C1-C2} are met, we immediately achieve the goal \eqref{eq8}.

\end{document}